\documentclass[aps,prl,nofootinbib,reprint,superscriptaddress,twocolumn,amsmath,amssymb]{revtex4-1}

\usepackage{graphicx}
\usepackage{dcolumn}
\usepackage{bm}
\usepackage{units}
\usepackage{microtype}
\usepackage{hyperref}
\usepackage{upgreek}

\begin{document}
\title{Mode-selective vibrational control of charge transport in $\pi$-conjugated molecular materials}

\author{Artem A.\ Bakulin*}
\email{aab58@cam.ac.uk}
\affiliation{FOM Institute AMOLF, Science Park 104, 1098 XG Amsterdam, The Netherlands}
\affiliation{Cavendish Laboratory, Univ. of Cambridge, JJ Thomson Avenue, Cambridge CB3OHE, UK}
\author{Robert Lovrin\v{c}i\'{c}*}
\affiliation{Department of Materials and Interfaces, Weizmann Institute of Science, Rehovot, Israel}
\affiliation{InnovationLab GmbH, 69115 Heidelberg, Germany}
\affiliation{Institut f\"{u}r Hochfrequenztechnik, TU Braunschweig, 38106 Braunschweig, Germany}
\author{Yu Xi*}
\affiliation{Department of Materials and Interfaces, Weizmann Institute of Science, Rehovot, Israel}
\author{Oleg Selig}
\affiliation{FOM Institute AMOLF, Science Park 104, 1098 XG Amsterdam, The Netherlands}
\author{Huib J. Bakker}
\affiliation{FOM Institute AMOLF, Science Park 104, 1098 XG Amsterdam, The Netherlands}
\author{Yves L.A. Rezus}
\affiliation{FOM Institute AMOLF, Science Park 104, 1098 XG Amsterdam, The Netherlands}
\author{Pabitra K. Nayak}
\affiliation{Department of Materials and Interfaces, Weizmann Institute of Science, Rehovot, Israel}
\author{Alexandr Fonari}
\affiliation{School of Chemistry and Biochemistry \& Center for Organic Photonics and Electronics, 
Georgia Institute of Technology, 901 Atlantic Drive NW, Atlanta, GA 30332-0400, USA}
\author{Veaceslav Coropceanu}
\affiliation{School of Chemistry and Biochemistry \& Center for Organic Photonics and Electronics, 
Georgia Institute of Technology, 901 Atlantic Drive NW, Atlanta, GA 30332-0400, USA}
\author{Jean-Luc Br\'{e}das}
\affiliation{School of Chemistry and Biochemistry \& Center for Organic Photonics and Electronics, 
Georgia Institute of Technology, 901 Atlantic Drive NW, Atlanta, GA 30332-0400, USA}
\affiliation{Solar \& Photovoltaics Eng.\ Res.\ Center, King Abdullah Univ.\ of Science and Technology, Thuwal 23955-6900, Kingdom of Saudi Arabia}
\author{David Cahen}
\email{david.cahen@weizmann.ac.il}
\affiliation{Department of Materials and Interfaces, Weizmann Institute of Science, Rehovot, Israel}

\begin{abstract}
*A.A.B., R.L., and Y.X.\ contributed equally.\\
\\
The soft character of organic materials leads to strong coupling between molecular nuclear and electronic dynamics. This coupling opens the way to control charge transport in organic electronic devices by exciting molecular vibrational motions. However, despite encouraging theoretical predictions, experimental realization of such control has remained elusive. Here we demonstrate experimentally that photoconductivity in a model organic optoelectronic device can be controlled by the selective excitation of molecular vibrations. Using an ultrafast infrared laser source to create a coherent superposition of vibrational motions in a pentacene/C60 photoresistor, we observe that excitation of certain modes in the $\unit[1500-1700]{cm^{-1}}$ region leads to photocurrent enhancement. Excited vibrations affect predominantly trapped carriers. The effect depends on the nature of the vibration and its mode-specific character can be well described by the vibrational modulation of intermolecular electronic couplings. Vibrational control thus presents a new tool for studying electron-phonon coupling and charge dynamics in (bio)molecular materials.\\
\textbf{Submitted}: Nov. 11, 2014\\
\textbf{Revised}: Feb. 3, 2015
\end{abstract}

\maketitle
The soft character of organic materials strongly influences their electronic functionality. \cite{bredas_polarons_1985,galperin_nuclear_2008} In these systems charge hopping and electronic delocalization are determined by the overlap of the molecular orbitals and, therefore, is highly sensitive to minor changes in molecular geometry. Hence, the electronic properties of organic materials are largely determined by the interplay between the electronic and nuclear dynamics of the molecules, referred to as vibronic coupling phenomena. A growing number of interdisciplinary studies show that vibronic effects lie at the heart of a diverse class of effects in physics, chemistry and biology - from non-linear behavior of molecular junctions \cite{galperin_nuclear_2008} to photophysics of vision \cite{polli_conical_2010} and even olfactory reception. \cite{gane_molecular_2013} Vibrational motions have been postulated to regulate the interaction between different molecular electronic states by modulating inter- and intra-molecular couplings, by donating or accepting extra energy quanta \cite{gane_molecular_2013,falke_coherent_2014}, and by suppressing \cite{ballmann_experimental_2012} or promoting \cite{chin_role_2013} quantum interference phenomena.

Vibronic effects were also shown to be fundamentally important for the conductivity of organic materials. Vibrational motions influence intermolecular electron tunneling probabilities \cite{pascual_selectivity_2003,asadi_polaron_2013,eggeman_measurement_2013} and govern a variety of non-equilibrium phenomena such as local heating \cite{ward_vibrational_2011}, switching \cite{galperin_nuclear_2008}, hysteresis, and electronic decoherence \cite{ballmann_experimental_2012,guedon_observation_2012}. This makes vibrational excitation a promising tool for spectroscopy of molecular junctions \cite{ward_vibrational_2011,selzer_transient_2013}, tracking charge transfer processes in organic- and bio-electronic systems, and, more generally, for the development of electronic devices. For example, remarkable opportunities for organic electronics would arise from the possibility to control charge transport, and, thus, affect device performance by coherently driving nuclear motions along a pre-selected reaction coordinate trajectory. However, despite many encouraging theoretical predictions \cite{sanchez-carrera_interaction_2010,ghezzi_hybrid_2011,wang_charge_2010} the experimental realization of vibrationally controlled electronics is still elusive due to the complexity of selective control of nuclear motions in an actual electronic junction.

Until now, vibration-associated charge dynamics in organic electronic devices have been only controlled with approaches that do not include mode selectivity. For example, the density and the equilibrium population of vibrational states have been varied via chemical synthesis of  molecules with different bond structures \cite{guedon_observation_2012} and via thermal population of low-frequency vibrations. \cite{ballmann_experimental_2012} However, in principle, it should be possible to access particular non-equilibrium nuclear or vibronic states by using instrumentation of optical time-resolved techniques, like visible pump-probe \cite{polli_conical_2010,falke_coherent_2014,ruban_identification_2007},  time-resolved stimulated/impulsive Raman \cite{kukura_femtosecond_2007,kraack_mapping_2013}, or transient IR absorption \cite{barbour_watching_2007}. For example, for inorganic perovskite materials \cite{rini_control_2007} and molecular Mott insulators \cite{kaiser_optical_2014} it has been reported that selective IR excitation can lead to strong modulation of the electronic properties by inducing a lattice phase transition. Sophisticated all-optical two-dimensional (2D) photon echo techniques are even capable of guiding a molecular system through a desired quantum superposition of vibronic/vibrational states \cite{chin_role_2013,brixner_two-dimensional_2005,tiwari_energy_2014,collini_coherently_2010,helbing_compact_2011}. Although such spectroscopic methods provide a comprehensive approach for probing and controlling molecular motions and have been applied to model systems such as molecular thin films or solutions, they have not yet been employed to control functional electronic (nano)devices.

In this work, we combine device characterization and ultrafast-spectroscopy methods to experimentally demonstrate that the performance of an organic optoelectronic system can be controlled by selectively exciting vibrational modes of the molecules involved in charge transport. As model system we use pentacene/C60 bi-layer photoresistors. Our experimental approach is based on the interferometric extension of the pump-push photocurrent (PPP) technique. \cite{bakulin_role_2012} In a PPP experiment, an optoelectronic device is illuminated by a sequence of laser pulses interacting with the active material in the device. The result of these interactions is detected by observing the variations in the current flow through the device as a function of time delay $T$ between the pump and push pulses and their spectra. Thus, PPP combines the sensitivity and device relevance of electronic methods with the excitation selectivity and ultrafast time resolution of optical methods. Since its introduction \cite{lukin_photoexcitation_1981,frankevich_formation_2000}, PPP has been applied and discussed in the context of photovoltaics \cite{bakulin_role_2012,muller_ultrafast_2005}, nanoelectronics, spectroscopy \cite{nardin_multidimensional_2013}, microscopy \cite{cocker_ultrafast_2013}, and molecular junction research \cite{selzer_transient_2013}. In this work, we extend the PPP method, using the recent progress in ultrafast interferometry \cite{helbing_compact_2011,skoff_simplified_2013} that allows for a precise control over the time/frequency-domain structure of the IR optical pulses. We apply a sequence of ultrafast mid-IR laser pulses to create a coherent superposition of molecular vibrational motions inside the active layer of a device and correlate this excitation with the device performance.

\begin{figure}
\includegraphics[width=\columnwidth]{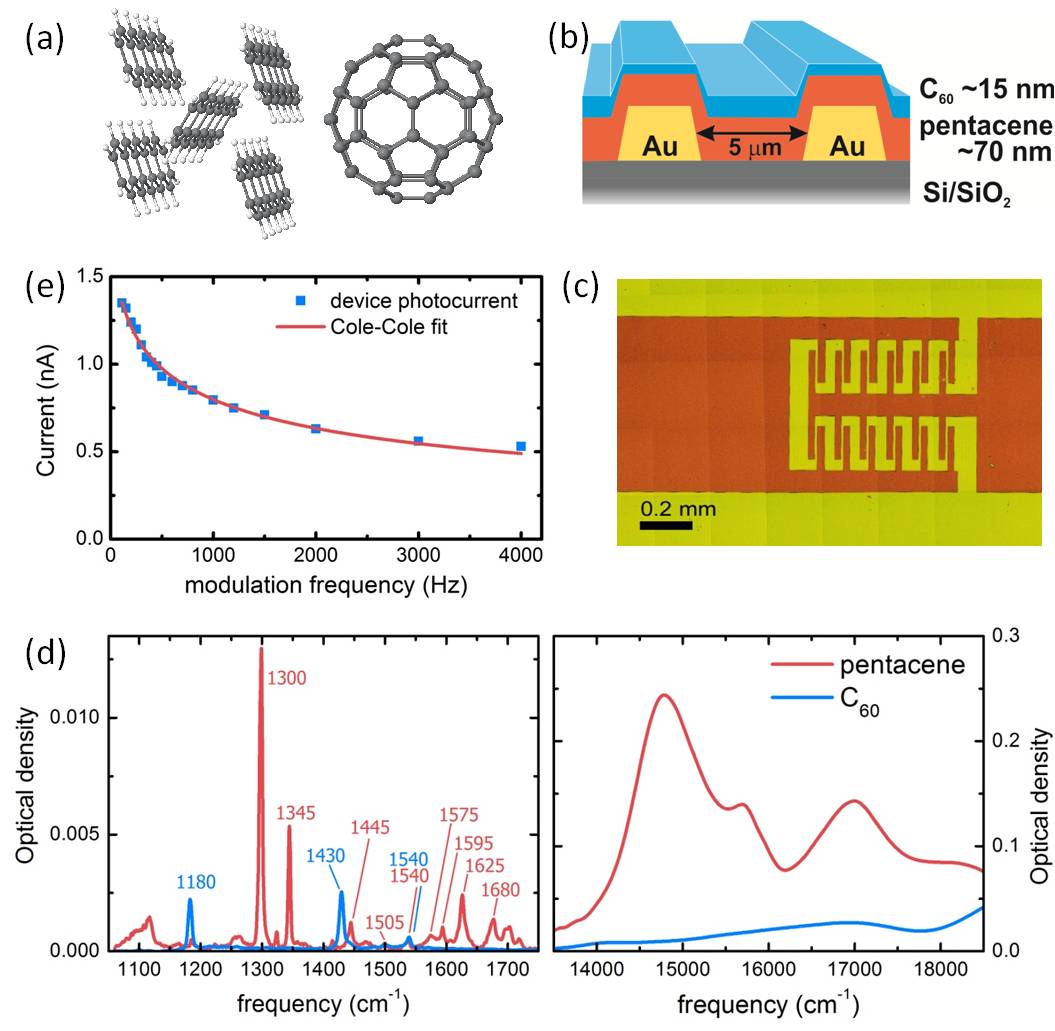}
\caption{\label{fig1} The molecular electronic device characterisation: (a) Molecular arrangement of molecules in the pentacene crystal and C60 fullerene structure. (b-c) Layout and microscope image of the device. (d) IR absorption in the vibrational fingerprint region and optical absorption spectra of pentacene and C60. The yellow shaded contour shows a typical laser spectrum used for IR push. (e) Photocurrent from the device as a function of visible light modulation frequency; the line is a Cole-Cole fit with a $\unit[2.2]{ms}$ lifetime constant and $\alpha=0.55$ dispersion parameter.}
\end{figure}

Figure \ref{fig1} a)-c) describes the organic bilayer photoresistor model system. The active layer of the device consists of polycrystalline pentacene ($\unit[70]{nm}$) and fullerene C60 ($\unit[15]{nm}$) films (Fig.\ \ref{fig1} a)), thermally evaporated on top of 3-, 5- or $\unit[10]{\upmu m}$ spaced electrodes arranged in a comb-like geometry on a SiO$_2$ substrate (Fig.\ \ref{fig1} b)). We chose this geometry rather than a sandwich-like structure, typical for photodiodes or solar cells, to improve the access of mid-IR pump-pulses to the active layer. Adding the C60 layer was critical to enhance the photocarrier generation in the film. \cite{wilson_singlet_2013}

Figure \ref{fig1} d) compares the absorption spectra of pentacene and C60 in the IR vibrational fingerprint region and in the region of the optical electronic transitions. C60 shows several distinct vibrational modes at $\unit[1180]{cm^{-1}}$, $\unit[1430]{cm^{-1}}$, and $\unit[1540]{cm^{-1}}$ and has a comparably low optical density in the visible. Pentacene has a rich spectrum of vibrational lines in the IR and also shows strong excitonic absorption features at frequencies above $\unit[14500]{cm^{-1}}$ ($\unit[690]{nm}$). According to density functional theory calculations, the strong IR peaks at 1300 and $\unit[1345]{cm^{-1}}$ are mostly associated with $\mathrm{C{=}C}$ stretching vibrations along the short axis of pentacene, while the weaker high-frequency vibrations correspond to atomic motions mostly aligned with the long axis of the molecule (see SI).

The dark I-V curves of the devices are symmetric and roughly linear, indicating good hole injection from the gold electrodes to the pentacene layer (see SI). Upon exposure to visible light, the current flow through the devices strongly increases ($\sim3$ times under $\unit[10]{mW/cm^2}$ illumination). Devices without a C60 layer demonstrated only negligible photoconductivity, which indicates that singlet (and triplet) \cite{wilson_singlet_2013} excitons generated after pentacene excitation are dissociating at the pentacene/C60 interface and that the charge generation proceeds through the interfacial charge transfer states \cite{deibel_role_2010}. Due to the large electron injection barrier at the pentacene/Au interface, the dark current is mostly provided by holes, while under illumination both holes and electrons contribute to the photocurrent. Unlike in a typical solar cell, both electrodes are placed below the pentacene films. Therefore, electrons and holes have to pass through the pentacene, which is known to lead to extremely long (up to seconds) extraction times of electrons residing in low-lying trap states in pentacene. \cite{gu_reversible_2007} This notion is confirmed by the dependence of the photocurrent on the light-modulation frequency (Fig.\ \ref{fig1} e)). Cole-Cole analysis of this dependence shows a typical time constant $>\unit[2]{ms}$, which we interpret as the lifetime of long-lived electronic charge carriers.

\begin{figure*}
\includegraphics[scale=0.5]{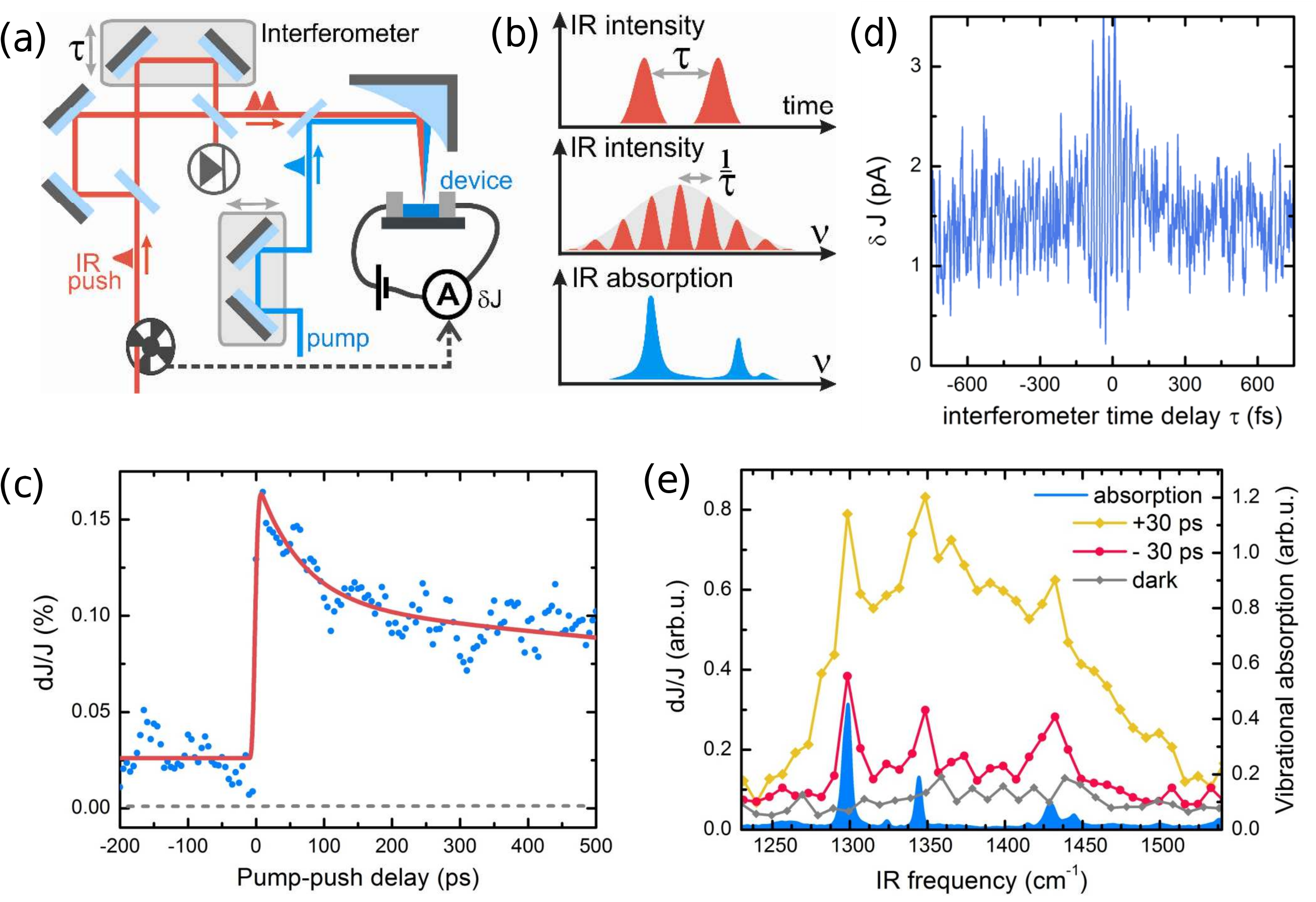}
\caption{\label{fig2} The results of pump-push photocurrent measurements: (a) The layout of the time- and frequency-resolved PPP experiment. (b) Time- and frequency-domain representations of the IR interferometric pulse pair, matching molecular vibrations. (c) Broadband PPP transient for pentacene/C60 photoresistor. (d) Typical PPP interferogram. (e) Frequency-resolved PPP signals for pentacene/C60 photoresistor, measured at different pump-push delay times $T$, and a no-pump dark measurement.}
\end{figure*}

Figure \ref{fig2} a) shows the layout of the experiment, designed to observe the effect of molecular vibrations on the charge transport through the device. The setup combines a $\unit[1]{kHz}$ visible-infrared ultrafast spectrometer and a lock-in current probe station wired to the device under $\sim \unit[5]{V}$ external bias. First, a visible ($\unit[15000]{cm^{-1}}$; $\unit[665]{nm}$; $\unit[1.9]{eV}$) pump pulse illuminates the device. The absorption of the pump light in pentacene leads to the build-up of excitons and charge carriers in the active layer. The generated carriers produce a sequence of $\unit[1]{ms}$-spaced current pulses in the measurement circuit with an average photocurrent $J\approx \unit[10]{nA}$, detected by the lock-in amplifier at $\unit[1]{kHz}$. We note that at such low current densities a charge-induced phase transition \cite{ando_evidence_2015} can be excluded. The device is irradiated with a push pulse at certain delay times $T$ before or after the pump pulse. The push pulse can promote $<1\%$ of the molecules to the excited vibrational state and can also excite low-frequency charge-associated IR electronic transitions. \cite{osterbacka_two-dimensional_2000} The effect of IR light on the charge separation and transport was detected via the variation of device photocurrent $\delta J$.

Figure \ref{fig2} c) presents a typical PPP transient, measured with a single-pulse push (one interferometer arm blocked) at $\unit[1250-1500]{cm^{-1}}$. When the pump was blocked we observed no signal due to the push only. At negative delay time $T$, when the push pulse arrives before the pump, we already observe a substantial increase of the current due to IR excitation (i.e., $\delta J>0$). We associate this response with the excitation of long-lived photocarriers that were generated by the preceding pump pulse that arrives $\sim \unit[1]{ms}$ earlier. This observation is in line with the long collection times of trapped carriers observed for electrons in pentacene. \cite{gu_reversible_2007} At delay time $T=0$, the PPP response promptly increases as the concentration of charges in the cell rises due to the arrival of the new pump pulse and the IR push influences their dynamics. The rapid rise is followed by a $\sim \unit[100]{ps}$ decay component that we assign to the geminate recombination of newly generated charge pairs, which are likely to form electrostatically bound charge-transfer excitons. \cite{bakulin_role_2012}

In a broadband experiment using a single-pulse push, it is not possible to distinguish the effects of low-frequency electronic excitations from the vibronic phenomena associated with the interference between the molecular vibrational motions and charge dynamics. To separate and address these phenomena individually, we performed push-frequency resolved measurements by exploiting the ultrafast interferometry approach. \cite{helbing_compact_2011} Using a Mach-Zehnder scheme (Fig.\ \ref{fig2} a)), the push beam is split into two pulses displaced in time by an interferometric delay $\tau$. This leads to the formation of a 1/$\tau$ periodic modulation in the total push spectrum (Fig. 2b) which allows for selective excitation of different coherent superpositions of modes within the bandwidth of the IR light. In a typical experiment, for a certain pump-push delay $T$, the signal $\delta J/J$ is detected as a function of interferometric delay $\tau$ (Fig. \ref{fig2} d)). The obtained interferogram is Fourier-transformed along the $\tau$ axis to yield the action spectrum of the push effect.

Figure \ref{fig2} e) shows a typical frequency-resolved PPP response of a pentacene/C60 device at negative and positive pump-push delay times $T$, and with no pump (dark). At both delays the response consists of a number of narrow peaks on top of a broad featureless response, roughly following the IR source spectrum. We associate the broad feature with intraband electronic and polaronic absorption, which typically spreads between 1000 and $\unit[5000]{cm^{-1}}$. \cite{osterbacka_two-dimensional_2000} The intraband excitation brings the associated charge carriers to a higher-lying delocalized state, thereby enhancing their mobility, decreasing their recombination, and thus increasing the current output. \cite{bakulin_role_2012} The narrow features in the PPP signal match well with the absorption peaks of the vibrational modes of pentacene and C60. Therefore these features in the frequency-resolved PPP response are assigned to the excitation of molecular vibrations that modulate the electronic dynamics. Interestingly, the broad electronic response dominates the PPP signal when the push arrives after the pump, while the vibrational features have similar amplitudes (within the experimental accuracy) at positive and negative $T$ delays. This observation indicates that the IR electronic excitation substantially promotes charge separation at the pentacene/C60 interface, soon after exciton generation. At the same time, the effect of vibrational excitation is present for long-lived trapped charge carriers and, therefore, does not influence charge separation, but only carrier de-trapping dynamics. \cite{bakulin_role_2012}

We now focus on the analysis of the vibrational features only. The effect of broadband electronic IR excitation on charge dynamics in organic semiconductors has been investigated previously \cite{bakulin_role_2012}, and is outside the scope of this paper. To study the effect of vibrational excitation for a broader set of vibrational modes, we use a wide push spectral window of $\unit[1150-1700]{cm^{-1}}$ and long $\tau$-scanning to obtain high frequency resolution. We also applied time-domain filtering (see SI) to suppress broad features due to electronic excitation and non-linear field-induced tunneling currents.

\begin{figure}
\includegraphics[width=\columnwidth]{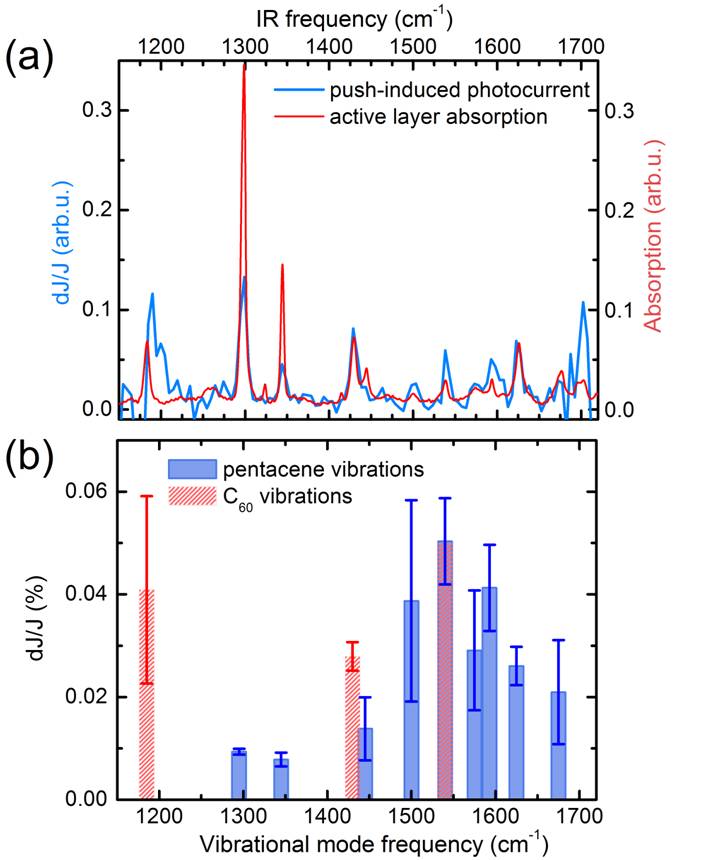}
\caption{\label{fig3} Experimental evaluation of vibrational control effect: (a) The vibrational part of the PPP response, measured at negative delay time ($\unit[-100]{ps}$). The signal amplitude is normalized to the spectral density of the IR push source. The spectrum is obtained using two PPP spectra, each covering a different but overlapping wavenumber range; these are scaled to match the amplitude of the $\unit[1430]{cm^{-1}}$ mode that is present in both spectra. For comparison, the absorption spectrum of the pentacene/C60 layer is presented in red. (b) The influence of different vibrations on device photocurrent, estimated by normalizing the amplitude of the PPP signal to the absorbed IR intensity. The change of photocurrent absolute value corresponds to a flat $\unit[0.5]{J/cm^2}$ per cm$^{-1}$ spectral density of exciting IR light, fully absorbed by the vibrations. The error bars are calculated from standard deviations for measurements on different devices; the number of measurements was 10 for $\unit[1300-1450]{cm^{-1}}$ modes and 4 for all other modes.}
\end{figure}

Figure \ref{fig3} a) presents the vibration-associated PPP spectrum covering most of the IR fingerprint frequency range. The amplitude of the PPP response was normalized to the spectral density of the push pulse to allow for a direct comparison of the different vibrational lines. The spectrum is the result of several measurements with different push frequencies spliced together to match the amplitude of the effect for the $\unit[1430]{cm^{-1}}$ feature, which was present in all measurements.  In the $\unit[1150-1700]{cm^{-1}}$ region, we observe twelve PPP peaks at frequencies that match well the IR-active vibrational modes of pentacene and C60.

We note that, while the vibrations of charged pentacene may differ from those of the neutral molecules \cite{capek_dynamics_1995}, these differences should not be observed in the PPP data. Firstly, the shift in frequency for most individual modes is small \cite{brinkmann_electronic_2004,cazayous_iodine_2004} and, for most modes, below our frequency resolution ($\sim \unit[10]{cm^{-1}}$). This conclusion is also supported by DFT calculations, see Suppl. Information, Tables S1 and S2. Secondly, the minor shifts of the vibrational levels lead to a highly efficient vibrational energy transfer \cite{woutersen_resonant_1999} between neutral and charged molecules, which allows the vibrational excitation of neutral pentacene to be delivered to the charge trapping sites. Thirdly, according to Miller-Abrahams formalism, when a carrier hops from a radical to a neutral state all vibrational modes coupled to these electronic states contribute to the transfer rate. \cite{miller_impurity_1960} Therefore, it is not surprising that the lattice vibrations, i.e. those of the neutral pentacene, are observed in the detrapping dynamics.

We observe that the amplitude of the PPP response does not follow the intensity of the IR absorption. For example, the band at $\unit[1345]{cm^{-1}}$ possesses a much stronger IR absorption than the $\unit[1630]{cm^{-1}}$ vibration, but shows a weaker PPP response. This result shows that the observed PPP response cannot be explained by the equilibration of vibrational energy between modes and average heating of the device active layer, thus illustrating the mode-selective character of the PPP response. This example illustrates that different atomic motions couple differently to the charge dynamics of the system. To exclude that the non-scaling of the PPP response with IR absorption is merely an effect of a different orientation of the vibration dipoles with respect to the exciting IR light, we also performed angle-dependent IR absorption measurements. These measurements showed that the modes exhibiting very different PPP effect, e.g. at $\unit[1345]{cm^{-1}}$ and $\unit[1630]{cm^{-1}}$, have similar dipole orientations (see SI), which rules out orientation effects.

Figure \ref{fig3} b) compares the effect of vibrational excitation on the device photoconductivity for different vibrational modes, obtained by normalizing the PPP response to the number of photons absorbed by the vibrational mode. In accordance with Figure 3a, the 1300 and $\unit[1345]{cm^{-1}}$ modes show the weakest coupling. The higher-frequency vibrations of pentacene show a 5-8 times higher effect on the photoconductivity. For two of the fullerene vibrations the effect is similar to that of the high-frequency vibrational modes of pentacene.

These results can be rationalized in the framework of the phonon-assisted Miller-Abrahams (MA) theory. \cite{miller_impurity_1960} According to this model, carrier hopping from a trapping state to higher-energy (more conducting) states takes place via absorption of a phonon with energy $\Delta$ to compensate for the energy difference between initial and final electronic states; the hopping rate $k$ is defined by the electron-vibrational coupling constant ($\nu$) and the occupation number ($n_{\Delta}$) of the absorbed phonon, i.e.
\begin{equation}
k \propto \nu^2 n_{\Delta}.
\end{equation}
In thermal equilibrium, the occupation number of a high-energy molecular vibration is very small:
\begin{equation}
n_{\Delta}=\exp{(-\Delta/k_{\mathrm{B}}T)}.
\end{equation}
For a comprehensive description of the PPP response the interaction with the IR photons should be included into the MA model. However, at the conceptual level the effect can be understood by assuming that an IR excitation creates a non-equilibrium population of the molecular vibrational manifold; therefore, an increase in the hopping probability is expected. The mode-selective character of the PPP response is therefore defined by the electron-vibration coupling constants.

\begin{figure*}
\includegraphics[scale=0.65]{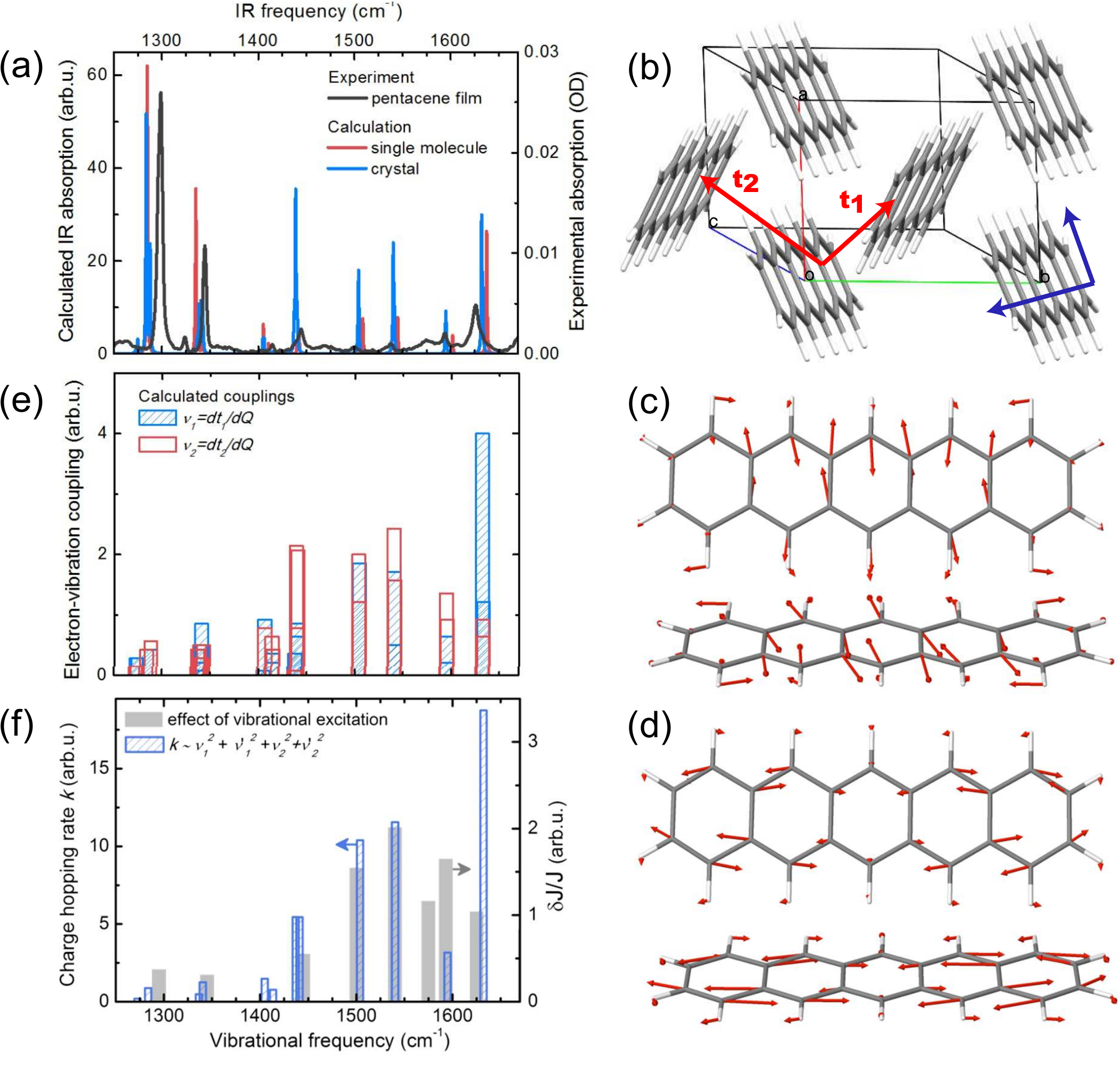}
\caption{\label{fig4} (a) Simulated IR spectrum for the pentacene molecule (red) and crystal (blue), superimposed on the experimental spectrum for comparison. (b)Molecular crystal structure and transfer integrals ti addressed in the calculations. (c,d) Eigendisplacements for the modes with large transition dipoles at $\unit[1288]{cm^{-1}}$ and $\unit[1632]{cm^{-1}}$. (e) Non-local electron-phonon coupling constants for two largest transfer integrals. (f) Calculated vibration-induced hopping rates for the different modes that were addressed experimentally, compared to the experimentally observed effect of vibrational excitation on the photocurrent, $\delta J/J$.}
\end{figure*}

To achieve mechanistic insight into the observed phenomena we performed a theoretical analysis of the coupling between the different molecular vibrations and the charge carriers (holes) in pentacene. In molecular systems these couplings can be divided into two types, i.e. local (Holstein-type) and non-local (Peierls-type). \cite{bredas_organic_2002} The Holstein electron-phonon interaction originates from the modulation of the site energies by the vibrations. Only totally symmetric molecular vibration modes can contribute to this interaction. For centrosymmetric molecules like pentacene, the symmetric modes are not IR-active and will not absorb IR photons, so that we can rule out Holstein electron-phonon coupling effects in the PPP response. The Peierls-type electron-phonon couplings are associated with the dependence of the transfer integrals on the distances between adjacent molecules and their relative orientations. \cite{bredas_organic_2002} For this type of coupling, there are no symmetry restrictions.

Based on previous studies \cite{bouchoms_morphology_1999}, we used a triclinic polymorph \cite{campbell_crystal_1962} to represent the pentacene layer structure in the calculations. Figure \ref{fig4} a) shows the simulated IR spectra for a single pentacene molecule and for the crystal in comparison to the experimental IR absorption. The agreement between the experimental and calculated vibrational frequencies (with typical discrepancies $< \unit[10]{cm^{-1}}$) allows the assignment of the vibrational modes observed in the frequency-resolved PPP experiment. The 1300 and $\unit[1345]{cm^{-1}}$ features in the experimental spectrum are associated (Fig.\ \ref{fig4} c) and SI) with in-plane ring stretching modes along the short axis of pentacene \cite{hosoi_infrared_2005,salzmann_phase_2007}, while the IR peaks at $\unit[1540]{cm^{-1}}$ and $\unit[1630]{cm^{-1}}$ are associated with molecular deformations along the long axis of pentacene (Fig.\ \ref{fig4} d) and SI). 

The non-local hole-vibration couplings are defined as the derivatives of the charge transfer integrals with respect to the vibrational coordinates, $\nu_i=dt_i/dQ_j$, and can be computed numerically. \cite{sanchez-carrera_interaction_2010,coropceanu_interaction_2009} Both the transfer integrals and electron-vibration couplings have been derived in a one-electron approximation (see SI for details). Our results indicate that there are two main transfer integrals contributing to charge transfer in the pentacene crystal: $t_1=\unit[75]{meV}$ and $t_2=\unit[32]{meV}$; both are associated with intermolecular interactions along the herringbone directions (see the red and blue arrows in Fig.\ \ref{fig4} b). The other two transfer integrals oriented along the a-axis are substantially smaller and do not demonstrate substantial modulation by IR-active modes. (See SI) The derived coupling constants of the IR-active modes are shown in Figure 4e. The couplings in the $\unit[1400-1650]{cm^{-1}}$ range are about 2 to 5 times larger than in the $\unit[1200-1400]{cm^{-1}}$ range. To link the variations in coupling and the probability to de-trap a charge with a vibrational excitation, we estimated the rates of vibration-induced charge hopping. In the case of two pathways, the rate is defined within perturbation theory as:
\begin{equation}
k_i \propto (\nu_{1,i}^2+\nu_{2,i}^2+\nu_{1,i}'^2+\nu_{2,i}'^2),
\end{equation}
where $\nu_{j,i}$ and $\nu_{j,i}'$ correspond to quasi-degenerate molecular vibrations of similar frequencies. Figure \ref{fig4} f) presents these hopping rates together with the experimental observations from Figure \ref{fig3} b). The theoretical results are fully consistent with the experimental data. In particular, they capture well the mode-selective character of the phenomena, with the modes below $\unit[1430]{cm^{-1}}$ calculated to have a much smaller impact on charge hopping than the higher-frequency ones. Based on the calculations, the intermolecular electronic couplings and charge transport in pentacene crystals are seen to be most sensitive to stretching deformations along the long molecular axis, while the stretching deformations along the short molecular axis are less important.

In conclusion, we demonstrated that the vibrational coupling phenomena, which play an essential role in molecular-scale charge transport, can be explored and put to action by combining optical and electronic techniques. Both the experiment and theoretical calculations demonstrate that different non-equilibrium geometries and atomic motions have different effects on the charge dynamics. Specifically, our results show that vibrations along the long axis of pentacene molecules lead to a stronger increase of hopping transport via charge de-trapping than vibrations along the short axis. The mode-selective vibrational control of charge dynamics introduced here opens up a plethora of opportunities for basic research, including the development of high-mobility organic semiconductors, and the utilization of vibronic phenomena for ultrafast switching of organic devices. In addition, the mode-selective and local nature of our method might be particularly useful for the identification of charge transport mechanisms and pathways in (bio)molecular junctions. \cite{cordes_electron_2009}

\section*{Acknowledgements}
We thank Richard Friend, Ayelet Vilan and Dassia Egorova, as well as the reviewers, for useful discussions, and Johannes Zimmermann and Tobias Glaser for the angle-dependent IR spectra. This work was supported by the Netherlands Organization for Scientific Research Onderzoek (NWO) through the ``Stichting voor Fundamenteel Onderzoek der Materie'' (FOM) research program. A.A.B. also acknowledges a VENI grant from the NWO. A.A.B. is currently a Royal Society University Research Fellow. R.L.\ acknowledges a Marie Curie IE Fellowship from the EU, held at the Weizmann Institute. X.Y. thanks the Council for Higher Education (Israel) for a PBC program postdoctoral research fellowship. V.C.\ and J.L.B.\ thank support from the Office of Naval Research and MURI Center on Advanced Molecular Photovoltaics, award No. N00014-14-1-0580. D.C.\ thanks the Israel Science Foundation Centre of Excellence program, the Grand Centre for Sensors and Security and the Schmidt Minerva Centre for Supramolecular Architecture for partial support. D.C. holds the Sylvia and Rowland Schaefer Chair in Energy Research.

\end{document}